\RequirePackage{fix-cm}

\documentclass[twocolumn]{svjour3}     
\smartqed  
\usepackage{graphicx}
\usepackage{csquotes}
\usepackage[hidelinks]{hyperref}
\usepackage{cite}
\usepackage{amsmath,amssymb,amsfonts}
\usepackage{algorithmic}
\usepackage{textcomp}
\usepackage{framed,multirow}
\usepackage{float}
\usepackage{mathrsfs}
\usepackage{footnote}
\usepackage{xcolor}
\usepackage{multirow}
\usepackage{array}
\usepackage{tikz}
\usepackage{xurl}
\usepackage{dblfloatfix}
\newcolumntype{P}[1]{>{\raggedright\arraybackslash}p{#1}}

\begin{document}
	
	\title{RCA-IUnet: A residual cross-spatial attention guided inception U-Net model for tumor segmentation in breast ultrasound imaging}
	
	\titlerunning{RCA-IUnet for tumor segmentation in breast ultrasound imaging}        
	
	\author{Narinder Singh Punn*         \and
		Sonali Agarwal 
	}
	
	\authorrunning{Punn N S, Agarwal S} 
	
	\institute{Narinder Singh Punn* \at
		IIIT Allahabad, Prayagraj, India, 211015 \\
		Tel.: +91-7018466740\\
		\email{pse2017002@iiita.ac.in}           
		\and
		Sonali Agarwal \at
		IIIT Allahabad, Prayagraj, India, 211015 \\
	}
	

	\maketitle
	
	\begin{abstract}
		The advancements in deep learning technologies have produced immense contributions to biomedical image analysis applications. With breast cancer being the common deadliest disease among women, early detection is the key means to improve survivability. Medical imaging like ultrasound presents an excellent visual representation of the functioning of the organs; however, for any radiologist analysing such scans is challenging and time consuming which delays the diagnosis process. Although various deep learning based approaches are proposed that achieved promising results, the present article introduces an efficient residual cross-spatial attention guided inception U-Net (RCA-IUnet) model with minimal training parameters for tumor segmentation using breast ultrasound imaging to further improve the segmentation performance of varying tumor sizes. The RCA-IUnet model follows U-Net topology with residual inception depth-wise separable convolution and hybrid pooling (max pooling and spectral pooling) layers. In addition, cross-spatial attention filters are added to suppress the irrelevant features and focus on the target structure. The segmentation performance of the proposed model is validated on two publicly available datasets using standard segmentation evaluation metrics, where it outperformed the other state-of-the-art segmentation models. 
		\keywords{Breast tumor segmentation \and Deep learning \and Ultrasound imaging \and U-Net}
		
	\end{abstract}
	
	\section{Introduction}
	Breast cancer is the most prevalent cancer in women among all the cancers \cite{siegel2019cancer} with the leading cause of death worldwide. With the molecular etiology of breast cancer being unknown, identifying the early signs of cancer is the only means to reduce the mortality rate. Due to the non-invasive, non-radioactive, painless, cost effective and ease in availability of the ultrasound imaging \cite{cheng2010automated}, it is most widely accepted for screening and diagnosing breast cancer. However, even for an expert radiologist, the manual analysis of such scans is challenging and time consuming. Following this context, deep learning based computer-aided diagnosis (CAD) systems are developed for the early detection of breast tumor for faster diagnosis and treatment \cite{xian2018automatic}. In most CAD systems breast tumor segmentation (BTS) is the key phase for follow up tumor treatment plans and diagnosis, where the goal is to segregate the target tumor region from the rest of the image. However, most of the approaches proposed for BTS are presented and validated on the private datasets which limit their reusability and reachability.
	
\begin{figure} []
		\centering
		\includegraphics[width=\columnwidth] {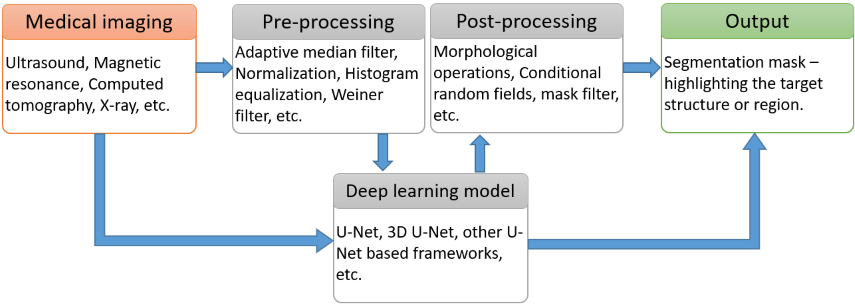}
		\caption{Generalized representation of the overview of the biomedical image segmentation models.}
		\label{fig1}
	\end{figure}

	The general schematic representation of the deep learning based segmentation models is presented in Fig. \ref{fig1}. In the data pre-processing phase, the aim is to transform the data into the trainable format by applying certain techniques like normalization to reduce intensity variation, resize to fit the model input layer, cropping the irrelevant features or noise, data augmentation, etc. The processed data is utilized to train the deep learning model and generate the desired segmentation mask. Finally, the generated mask is post-processed to refine the segmentation results. In the last decades many deep learning based segmentation models are proposed \cite{haque2020deep}, where U-Net based approaches achieved state-of-the-art performance in a wide variety of 2D and 3D data space \cite{punn2020inception,oktay2018attention,dong2020deu} while also addressing the challenge of limited availability of the medical data.
	
	\subsection{U-Net}
	The U-Net model, developed by Ronneberger et al. \cite{ronneberger2015u}, formed the basis of the state-of-the-art biomedical image segmentation networks. This model employed unique contraction and expansion paths along with the residual skip connections for biomedical image segmentation. In this architecture, the contraction phase tends to extract high and low level features, whereas the expansion phase follows from the features learned in the corresponding contraction phase (skip connections) to reconstruct the image into the desired dimensions with the help of transposed convolutions or upsampling operations. The network does not have any fully connected layers and only uses the valid convolution accompanied by rectified linear unit (ReLU) activation and max pooling operations. Following the state-of-the-art potential of the U-Net model, many variants are proposed for biomedical image segmentation \cite{haque2020deep}. With such high utility of the U-Net model, this article presents a U-Net based model for breast tumor segmentation.
	
	\subsection{Our contribution}
	The major contribution of the article concerning breast tumor segmentation is described below:
	\begin{itemize}
		\item A novel architecture, residual cross-spatial attention guided inception U-Net model (RCA-IUnet) is introduced with long and short skip connections to generate binary segmentation mask of tumor using ultrasound imaging.
		\item Instead of the direct concatenation of encoder feature maps with upsampled decoded feature maps, a cross-spatial attention filter is introduced in the long skip connections that use multi-level encoded feature maps to generate attention maps for concatenation with decoded feature maps.
		\item Hybrid pooling operation is introduced that uses a combination of spectral and max pooling for efficient pooling of the feature maps. It is utilized in two modes: a) same: used in inside inception block  b) valid: used to connect inception blocks (reduce the spatial resolution by half the input feature map).
		\item The model is also equipped with short skip connections (residual connections) along with the inception depth-wise separable convolution layers (concatenated feature maps from 1$\times$1, 3$\times$3, 5$\times$5 and hybrid pooling).
	\end{itemize}
	
	\subsection{Article organization}
	The rest of the article is structured in various sections covering related work in Section 2 to present the literature survey and the proposed approach in Section 3. In the later Section 4 and Section 5, the experimental setup and results are presented along with the qualitative and quantitative results to cover the comparative analysis and ablation study. Finally, the concluding remarks and future scope are presented.
	
	\section{Related work}
	With the advent of advancements in deep learning, the healthcare sector is improving every day \cite{Bhardwaj2017ASO}.  In classical approaches, thresholding \cite{shan2008novel}, region growing \cite{joo2004computer} and watershed \cite{huang2006automatic} based frameworks were adopted to produce segmentation masks. In this section, various breast ultrasound image segmentation approaches are studied that achieved state-of-the-art performance, especially on their private dataset \cite{xian2018automatic}.
	
	Shan et al. \cite{shan2012completely} proposed a fully automatic deep learning based segmentation framework to identify and localize the breast lesions using ultrasound imaging. The framework considers textural and spatial features, where initially region of interest (RoI) is generated (region likely to contain lesion) with automatic seed point selection and region growing approach. Following the RoI generation, multi-domain features are extracted: phase in max orientation (PMO), radial distance (RD) and a frequently used texture-and-intensity feature joint probability (JP). Later, an artificial neural network was used to generate the binary segmentation mask of the lesion region. In 2014, Torbati et al. \cite{torbati2014efficient} introduced a neural network based framework that uses merging moving average self organizing maps (MMA-SOM) to generate an initial segmentation mask and objects belonging to the joint cluster are merged. Later, a 2D discrete wavelet transform (DWT) is computed to generate the input feature space of the network. The approach was validated on multiple modalities, where for breast ultrasound image segmentation authors established a strong correlation between ground truth mask and predicted mask. In another approach, a stacked denoising auto-encoder (SDAE) was introduced by Cheng et al. \cite{cheng2016computer} to diagnose lesions in breast ultrasound and pulmonary nodules in CT scans. The approach achieved robust results and outperformed traditional computer-aided diagnosis (CAD) approaches, because of automatic feature extraction and high noise tolerance.
	
	With transfer learning \cite{tan2018survey} being a growing area of research, Huyanh et al. \cite{huynh2016mo} proposed a transfer learning based approach to classify cystic, benign, or malignant cancer in breast ultrasound imaging. In a similar approach, Fujioka et al. \cite{fujioka2019distinction} utilized GoogLeNet inception \cite{szegedy2015going} model to classify breast tumors with varying shapes and size. To generate the segmentation mask, Yap et al. \cite{yap2017automated} utilized a pre-trained FCN-AlexNet model. The approach outperformed other segmentation models, however failed to produce better segmentation masks for small lesion regions. Huang et al. \cite{huang2020segmentation} introduced a superpixels classification and clustering patches based segmentation approach to diagnose breast tumors in ultrasound imaging. Though the authors achieved promising segmentation results, the performance was fairly low on large tumors due to simple linear iterative clustering \cite{ilesanmi2020multiscale}. In order to generate better segmentation results, several methods have been studied to dynamically adapt to the target structures (tumor) of varying shapes and sizes using attention mechanism \cite{vaswani2017attention, oktay2018attention}. Following this context, Lee et al. \cite{lee2020channel} introduced a channel attention module and multi-scale grid average pooling to segment breast ultrasound images. Unlike channel attention that offers depth correlation, spatial attention allows to prioritize an area within the receptive field to better extract the target feature maps \cite{woo2018cbam}. With this potential of spatial attention filter, we propose a novel residual inception U-Net architecture that uses a cross-spatial attention filter to extract relevant features from multi-scale encoded features to generate binary tumor segmentation masks. Furthermore, the model is equipped with residual inception depth-wise separable convolution and hybrid pooling (max pooling and spectral pooling) layers for better feature extraction and learning.
	
	\begin{figure}[t]
		\centering
		\includegraphics[width=\columnwidth] {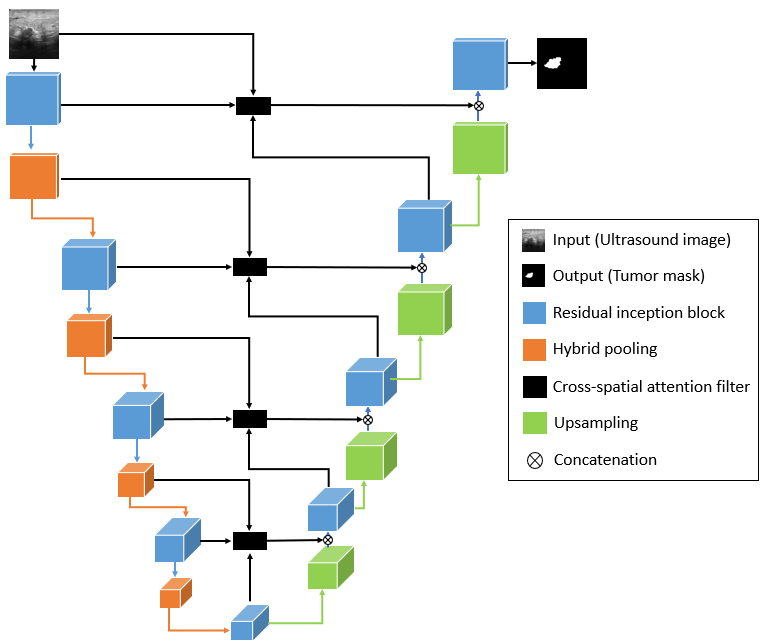}
		\caption{Schematic representation of the RCA-IUnet.}
		\label{fig4}
	\end{figure}
	
	\begin{figure*}[]
		\centering
		\includegraphics[scale=0.53] {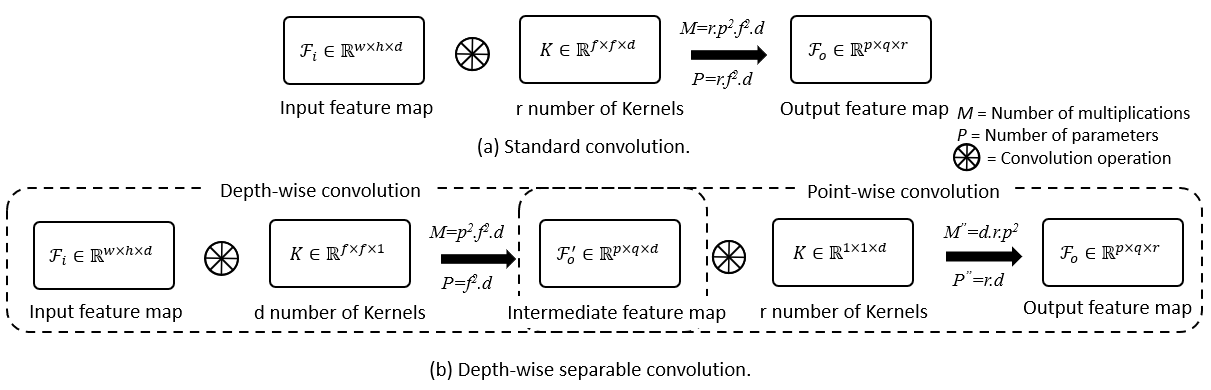}
		\caption{Convolution operations: a) Standard convolution, and b) Depthwise separable convolution.}
		\label{fig5}
	\end{figure*}
	
	\section{Proposed architecture}
	
	The schematic representation of the residual cross-spatial attention guided inception U-Net model (RCA-IUnet) is presented in Fig. \ref{fig4}. The network follows U-Net topology where standard convolution and pooling operations are replaced by inception convolution with short skip connections and hybrid pooling along with the cross-spatial attention filter on long skip connection to focus on the most relevant features. The network has four stages of encoding and decoding layer, where at each stage the spatial dimension (width and height) of the feature map reduces by 50\% and channel depth increases by 50\%. Besides, in order to minimize the training parameters and the number of multiplications, the depth-wise separable convolution (DSC) operation \cite{chollet2017xception} is followed which resulted in 2.9M trainable parameters. 
	
	The network generates a binary segmentation mask to highlight the tumor region. In some of the predicted masks, minor holes (false negative) and small unnecessary regions (false positive) are identified. Hence, the generated segmentation mask is further refined with post-processing morphological operations such as the flood fill algorithm, mask extraction and binary thresholding to fill the minor holes left in the generated mask based on the surrounding or connected pixels (reducing the false negative predictions), remove the small masked regions (reducing the false positive predictions) and filter the masked regions, respectively.
	
	\subsection{Depthwise separable convolution}
	Unlike standard convolution (SC) operation, in DSC the convolution is performed in two stages involving depthwise and pointwise convolutions as shown in Fig. \ref{fig5}(b) for some input feature map with width (w), height (h) and depth (d), $\mathcal{F}\in\mathbb{R}^{w\times h\times d}$ . From Fig. \ref{fig5} it can be observed that the ratio of reduction in parameters and multiplications can be presented using Eq. \ref{eq1} in terms of number of parameters ($P_{SC}, P_{DSC}$) or multiplications ($M_{SC}, M_{DSC}$), number of kernels ($r$) and kernel size ($f$).

	\begin{equation}
		M_{SC}=r.p^2.f^2.d \;,\;\; P_{SC}=r.f^2.d
		\label{eq1_1}
	\end{equation}
	
	\begin{equation}
		M_{DSC}=d.p^2.
		(f^2+r) \;,\;\; P_{DSC}=d.(f^2+r)
		\label{eq1_2}
	\end{equation}
	
	\begin{equation}
		\frac{M_{DSC}}{M_{SC}}=\frac{P_{DSC}}{P_{SC}}=\frac{1}{r}+\frac{1}{f^{2}}
		\label{eq1}
	\end{equation}
	
	\subsection{Hybrid pooling}
	In deep learning various pooling operations are introduced \cite{akhtar2020interpretation}, where max pooling is the most common choice for downsampling the feature maps. Max pooling tends to only preserve the sharpest features by applying max operation in given window size, whereas spectral pooling \cite{rippel2015spectral} not only downsamples the feature maps but also preserves more information as compared to max pooling. In spectral pooling, discrete Fourier transform (DFT) of the input feature map is computed to truncate the high frequency values in the spectral domain and then inverse DFT is applied to convert back to the spatial domain. Hence, to better downsample the feature maps, in this article hybrid pooling is introduced in which downsampled feature maps from max pooling and spectral pooling are merged using the 1$\times$1 convolution operation.	
	
	\subsection{Inception convolution}
\begin{figure}[]
		\centering
		\includegraphics[width=0.9\columnwidth] {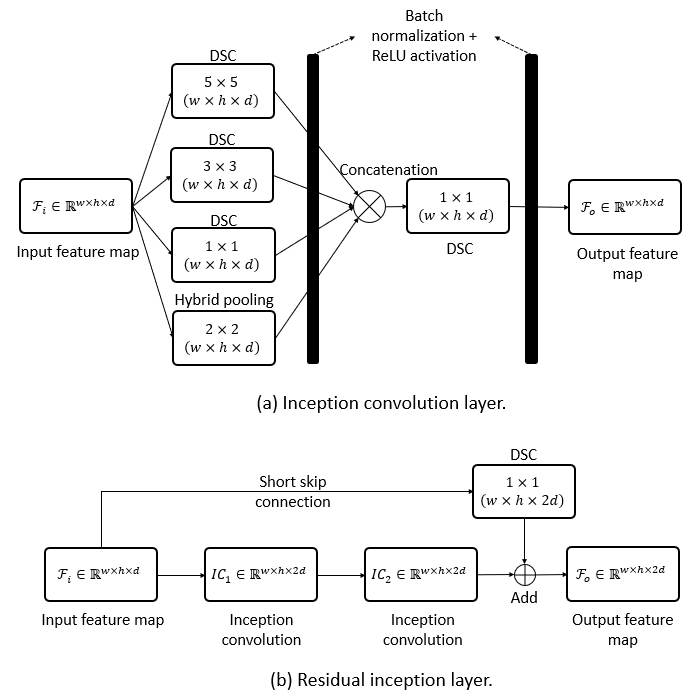}
		\caption{Overview of the a) Inception convolution layer and b) Residual inception layer.}
		\label{fig7}
	\end{figure}

	In order to identify the features concerning tumor regions of varying shape and size, the model needs to have an adaptive receptive field \cite{punn2020multi, luo2017understanding}. The inception convolution is designed by concatenating the feature maps extracted using the ReLU activated parallel depthwise separable convolutions with different kernels of sizes such as 1$\times$1, 3$\times$3 and 5$\times$5, and hybrid pooling while also using the batch normalization to avoid the covariance shift problem. Finally, the concatenated feature maps undergo 1$\times$1 convolution to setup the channel correlation and optimize the spatial dimension. Consider an input feature map, $\mathcal{F}_{i}\in\mathbb{R}^{w\times h\times d}$, the overview of the inception convolution is illustrated in Fig. \ref{fig7}(a). Following from the inception convolution layers the residual inception convolution block is developed by applying double inception convolution layers with a short skip connection to merge the extracted feature maps with input using 1$\times$1 DSC as shown in Fig. \ref{fig7}(b).
	
	\subsection{Cross-spatial attention block}
\begin{figure}[]
		\centering
		\includegraphics[width=\columnwidth] {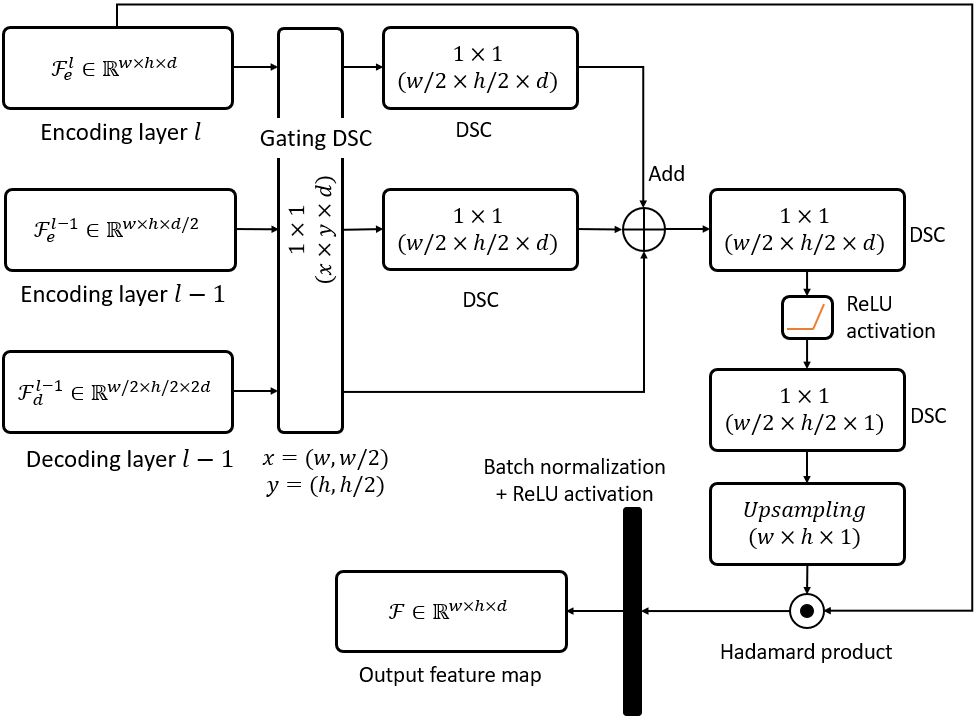}
		\caption{Schematic representation of cross-spatial attention block.}
		\label{fig8}
	\end{figure}

	In order to draw the attention of the model towards the tumor structure of varying shape and size, a cross-spatial attention block is introduced in the long skip connections. Unlike the standard attention network \cite{oktay2018attention}, in this block, the attention filter utilizes the extracted features maps from multiple encoded layers to develop better correlation in the spatial dimension of the feature maps. The schematic representation of the cross-spatial attention approach is illustrated in Fig. \ref{fig8}, where feature maps from three different layers are considered to form the attention feature maps (output feature maps) which are later concatenated with the corresponding decoded layer in the expansion or reconstruction phase. 
	
	\section{Experiment setup}
	In this section, details concerning the experimental environment and datasets are presented along with the obtained results and comparative analysis. Due to non-availability of the implementation of the existing breast ultrasound image segmentation models and a standard testing set, the proposed model is compared with other state-of-the-art segmentation models like SegNet\footnote{\url{https://github.com/lsh1994/keras-segmentation}\label{segnet}} \cite{badrinarayanan2017segnet}, U-Net\textsuperscript{\ref{segnet}} \cite{ronneberger2015u}, U-Net++\footnote{\url{https://github.com/kannyjyk/Nested-UNet}\label{unetpp}} \cite{zhou2018unet++}, attention U-Net\footnote{\url{https://github.com/ozan-oktay/Attention-Gated-Networks}\label{aunet}} \cite{oktay2018attention}, dense U-Net\footnote{\url{https://github.com/clguo/Dense_Unet_Keras}\label{dunet}} and deep layer aggregation (DLA)\textsuperscript{\ref{unetpp}} \cite{yu2018deep} while using vgg16 \cite{he2016deep} and resnet50 \cite{he2016deep} as backbone architectures.
	
	\begin{table*}[]
		\centering
		\caption{Tumor segmentation evaluation metrics in terms of number of true positive (TP), true negative (TN), false positive (FP) and false negative (FN), predicted mask ($\mathcal{P}$) and ground truth ($\mathcal{G}$), $\mathcal{H}(\mathcal{P},\mathcal{G})$ is the mean of directed $AHD$ from $\mathcal{P}$ to $\mathcal{G}$ and $\mathcal{G}$ to $\mathcal{P}$ with $d$ as euclidean distance, $N$ is the total number of pixels and $t$ is the prediction threshold.}
		\label{tab1}
		\begin{tabular}{|P{1.2in}|l|}
			\hline
			Metric & Expression\\
			\hline
			Accuracy & $Acc=\frac{(TP+TN)}{(TP+TN+FP+FN)}$\\ \hline
			Precision & $Pr=\frac{TP}{(TP+FP)}$\\ \hline
			Recall & $R=\frac{(TP)}{(TP+FN)}$\\ \hline			
			Dice coefficient & $DC=\frac{2 \times |\mathcal{P} \cap \mathcal{G}|}{|\mathcal{P}|+|\mathcal{G}|}=\frac{2TP}{2TP+FP+FN}$\\ \hline
			\multirow{2}{1.2in}{Mean intersection-over-union} & $mIoU=\frac{1}{10}\sum_{t}IoU_t;$\;\;\;$t+=0.5\leq 0.95$\\
			& $IoU=\frac{\mathcal{P} \cap \mathcal{G}}{\mathcal{P}\cup \mathcal{G}}=\frac{TP}{TP+FP+FN}$\\ \hline
			\multirow{2}{1.2in}{Average Hausdorff distance} & $AHD=\frac{1}{2}\left(\frac{\mathcal{H}(\mathcal{P},\mathcal{G})}{\mathcal{P}}+\frac{\mathcal{H}(\mathcal{G},\mathcal{P})}{\mathcal{G}}\right)$\\
			&\;\;\;\;\;\;\;\;\;\;$=\frac{1}{2}\left(\frac{1}{\mathcal{P}} \sum_{p\in \mathcal{P}} \min_{g\in \mathcal{G}} d(p,g) + \frac{1}{\mathcal{G}} \sum_{g\in \mathcal{G}} \min_{p\in \mathcal{P}} d(p,g)\right)$\\ \hline
			Mean absolute error & $MAE=\frac{|\mathcal{P}-\mathcal{G}|}{N}$\\
			\hline  
		\end{tabular}
	\end{table*}
	
	\subsection{Dataset description and setup}
\begin{figure}
		\centering
		\includegraphics[width=\columnwidth] {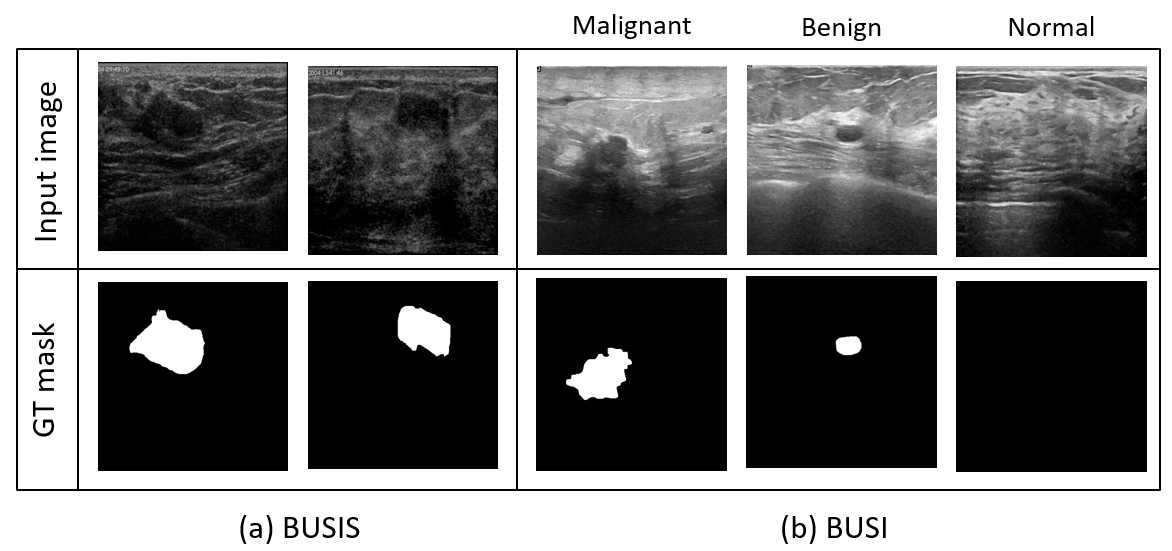}
		\caption{Breast ultrasound images with ground truth from a) BUSIS and b) BUSI datasets.}
		\label{fig9}
	\end{figure}

	The RCA-IUnet model is trained and evaluated using two publicly available datasets: a) breast ultrasound image segmentation (BUSIS) benchmark dataset \cite{xian2018benchmark} and b) breast ultrasound images (BUSI) dataset \cite{al2020dataset}. The BUSIS dataset comprises 562 breast ultrasound images that are collected from vivid hospitals: Harbin medical university, Qingdao university, and Hebei medical university. Each image is provided with a binary ground truth mask (1 label is assigned for tumor pixel and 0 label for background pixel) to highlight the tumor region which is generated using the majority voting approach from the annotations provided by various radiologists. Unlike the BUSIS dataset, BUSI dataset offers 780 ultrasound images divided into normal (133), benign (487) and malignant (210) classes along with the binary ground truth mask. Fig. \ref{fig9} shows the sample ultrasound images along with the ground truth from BUSIS and BUSI datasets. Due to the variation in the image size in both the datasets, the images are normalized and resized to 256$\times$256 for all the segmentation models. Both datasets are randomly split into 70\% of the training set and 30\% of the testing set and are kept the same throughout the experimentation. All the segmentation models are trained on the training set which is further split into 70\% train set and 30\% validation set. The trained models are then evaluated on the testing set.
	
	\subsection{Training and testing}
	The models are trained and tested on the BUSIS and BUSI datasets. The training phase is assisted with the stochastic gradient descent approach and Adam as an optimizer \cite{ruder2016overview} on an NVIDIA GeForce RTX 2070 Max-Q GPU. During training, the learning rate initialized at $1e-3$ is reduced by a factor of $2$ once learning stagnates to achieve better results. Moreover, earlystopping technique is adopted that halts the training process as soon as the validation error stops improving to avoid the overfitting problem. The RCA-IUnet is trained with the segmentation loss function ($\mathcal{L}$) that is defined as the average of binary cross entropy loss ($\mathcal{L}_{BC}$) and dice coefficient loss ($\mathcal{L}_{DC}$) as shown in Eq. \ref{eq2}.
	\begin{equation}
		\mathcal{L}=\frac{1}{2}    \mathcal{L}_{BC}+\frac{1}{2}\mathcal{L}_{DC}
		\label{eq2}
	\end{equation}
	\begin{equation}
		\begin{aligned}
			\mathcal{L}_{BC}\left(y,p\left(y\right)\right)=-\sum^N_i\left(y_i.{log \left(p\left(y_i\right)\right)}+\left(1-y_i\right).\right.
			\\
			\left.{log \left(1-p\left(y_i\right)\right)}\right)
		\end{aligned}
		\label{eq3}
	\end{equation}
	\begin{equation}
		\mathcal{L}_{DC}\left(y,p\left(y\right)\right)=1-\frac{2\sum^N_i{y_i.p(y_i)}}{\sum^N_i{{\left|y_i\right|}^{2}}\mathrm{+}\sum^N_i{{|{p(y}_i)|}^{2}}}
		\label{eq4}
	\end{equation}
	where $y$ is the ground truth label, $p(y)$ is the predicted label and $N$ is the total number of pixels. During the backpropagation the gradient of the loss function with respect to the predicted value can be computed using Eq. \ref{eq5}.
	
	\begin{equation}
		\frac{\partial \mathcal{L}}{\partial p(y)}=\frac{1}{2}\left[\frac{\partial \mathcal{L}_{BC}\left(y,p\left(y\right)\right)}{\partial p(y)}+\frac{\partial \mathcal{L}_{DC}\left(y,p\left(y\right)\right)}{\partial p(y)}\right]
		\label{eq5}
	\end{equation}
	where
	\begin{equation}
		\frac{\partial \mathcal{L}_{BC}\left(y,p\left(y\right)\right)}{\partial p(y)}=\frac{p\left(y\right) - y}{p\left(y\right)\left(1-p\left(y\right)\right)}
	\end{equation}
	\begin{equation}
		\frac{\partial \mathcal{L}_{DC}\left(y,p\left(y\right)\right)}{\partial p(y)}\mathrm{=-2}\left(\frac{y.({\left|y\right|}^2-{\left|p\left(y\right)\right|}^2}{{\left({\left|y\right|}^2+{\left|p\left(y\right)\right|}^2\right)}^2}\right)
	\end{equation}
	
	The trained models are utilized to predict the tumor segmentation mask for the test set. The performance of the models is compared using various evaluation metrics as shown in Table \ref{tab1}. In addition, inference time (IT)~\cite{amnon2020} is considered to measure the speed of the model. This is computed by measuring the average time taken by the model to generate mask for each sample in test set, where less inference time indicate faster mask generation.
	
	\begin{table*}[]
		\centering
		\caption{Comparative analysis of the RCA-IUnet with other segmentation approaches on the BUS datasets. The best results with post-processing (PP) no and yes are shown in bold blue and black fonts respectively.}
		\label{tab2}
		\resizebox{\linewidth}{!}{\begin{tabular}{|l|P{1in}|l|P{0.3in}|l|l|l|l|l|l|l|l|}
			\hline
			\multicolumn{1}{|l|}{\multirow{2}{*}{Dataset}} & \multirow{2}{*}{Model}                   & \multirow{2}{*}{Params} & \multirow{2}{0.3in}{IT (ms)} & \multirow{2}{*}{PP} & \multicolumn{7}{c|}{Tumor   segmentation}                                                                                                                                                                                                                  \\ 
			\cline{6-12}
			\multicolumn{1}{|l|}{}                         &                                          &                         &                                                                   &                     & Acc. $\uparrow$                   & Pr. $\uparrow$                    & R $\uparrow$                      & DC $\uparrow$                     & mIoU $\uparrow$                   & AHD $\downarrow$                  & MAE $\downarrow$                   \\ 
			\hline
			\multicolumn{1}{|c|}{\multirow{14}{*}{\rotatebox[origin=c]{90}{BUSIS}}}                       & \multirow{2}{1in}{SegNet + vgg16}          & \multirow{2}{*}{29.4M}  & \multirow{2}{*}{46.12}                                            & N                   & 0.910                             & 0.882                             & 0.713                             & 0.789                             & 0.777                             & 5.936                             & 0.052                              \\ 
			\cline{5-12}
			&                                          &                         &                                                                   & Y                   & 0.961                             & 0.949                             & 0.717                             & 0.817                             & 0.820                             & 5.730                             & 0.038                              \\ 
			\cline{2-12}
			& \multirow{2}{1in}{U-Net + resnet50}        & \multirow{2}{*}{36.5M}  & \multirow{2}{*}{25.48}                                            & N                   & 0.925                             & 0.899                             & 0.881                             & 0.890                             & 0.861                             & 5.303                             & 0.034                              \\ 
			\cline{5-12}
			&                                          &                         &                                                                   & Y                   & 0.980                             & 0.941                             & 0.889                             & 0.914                             & \textbf{0.910}                    & 4.800                             & 0.022                              \\ 
			\cline{2-12}
			& \multirow{2}{1in}{U-Net++ +  resnet50}     & \multirow{2}{*}{37.7M}  & \multirow{2}{*}{41.33}                                            & N                   & 0.928                             & 0.867                             & 0.847                             & 0.857                             & 0.888                             & 5.156                             & 0.029                              \\ 
			\cline{5-12}
			&                                          &                         &                                                                   & Y                   & 0.976                             & 0.921                             & 0.865                             & 0.892                             & 0.890                             & 4.920                             & 0.024                              \\ 
			\cline{2-12}
			& \multirow{2}{1in}{Attention U-Net + vgg16} & \multirow{2}{*}{31.9M}  & \multirow{2}{*}{45.32}                                            & N                   & 0.939                             & 0.894                             & 0.892                             & 0.893                             & 0.891                             & 4.980                             & 0.027                              \\ 
			\cline{5-12}
			&                                          &                         &                                                                   & Y                   & 0.978                             & 0.930                             & 0.888                             & 0.909                             & 0.909                             & 4.650                             & 0.022                              \\ 
			\cline{2-12}
			& \multirow{2}{1in}{Dense U-Net + vgg16}     & \multirow{2}{*}{20.2M}  & \multirow{2}{*}{41.54}                                            & N                   & 0.939                             & 0.838                             & 0.871                             & 0.854                             & 0.825                             & 5.485                             & 0.031                              \\ 
			\cline{5-12}
			&                                          &                         &                                                                   & Y                   & 0.973                             & 0.893                             & 0.872                             & 0.882                             & 0.879                             & 5.002                             & 0.027                              \\ 
			\cline{2-12}
			& \multirow{2}{1in}{DLA + vgg16}             & \multirow{2}{*}{23.7M}  & \multirow{2}{*}{44.22}                                            & N                   & 0.933                             & 0.840                             & 0.832                             & 0.836                             & 0.851                             & 5.345                             & 0.024                              \\ 
			\cline{5-12}
			&                                          &                         &                                                                   & Y                   & 0.976                             & 0.900                             & 0.834                             & 0.866                             & 0.887                             & 5.002                             & 0.023                              \\ 
			\cline{2-12}
			& \multirow{2}{1in}{RCA-IUnet (Ours)}        & \multirow{2}{*}{2.9M}   & \multirow{2}{*}{18.75}                                            & N                   & \textcolor{blue}{ \textbf{0.980}} & \textcolor{blue}{ \textbf{0.950}} & \textcolor{blue}{ \textbf{0.920}} & \textcolor{blue}{ \textbf{0.935}} & \textcolor{blue}{ \textbf{0.904}} & \textcolor{blue}{ \textbf{4.760}} & \textcolor{blue}{ \textbf{0.019}}  \\ 
			\cline{5-12}
			&                                          &                         &                                                                   & Y                   & \textbf{0.990}                    & \textbf{0.954}                    & \textbf{0.920}                    & \textbf{0.937}                    & \textbf{0.910}                    & \textbf{4.632}                    & \textbf{0.019}                     \\ 
			\hline
			\multicolumn{1}{|c|}{\multirow{14}{*}{\rotatebox[origin=c]{90}{BUSIS}}}                         & \multirow{2}{1in}{SegNet + vgg16}          & \multirow{2}{*}{29.4M}  & \multirow{2}{*}{46.15}                                            & N                   & 0.919                             & 0.779                             & 0.637                             & 0.701                             & 0.780                             & 5.896                             & 0.048                              \\ 
			\cline{5-12}
			&                                          &                         &                                                                   & Y                   & 0.925                             & 0.842                             & 0.693                             & 0.760                             & 0.787                             & 5.750                             & 0.042                              \\ 
			\cline{2-12}
			& \multirow{2}{1in}{U-Net + resnet50}        & \multirow{2}{*}{36.5M}  & \multirow{2}{*}{25.48}                                            & N                   & 0.825                             & 0.808                             & 0.815                             & 0.811                             & 0.815                             & 5.535                             & 0.035                              \\ 
			\cline{5-12}
			&                                          &                         &                                                                   & Y                   & 0.926                             & 0.881                             & 0.814                             & 0.846                             & 0.834                             & 5.050                             & 0.027                              \\ 
			\cline{2-12}
			& \multirow{2}{1in}{U-Net++ +  resnet50}     & \multirow{2}{*}{37.7M}  & \multirow{2}{*}{41.32}                                            & N                   & 0.885                             & 0.883                             & 0.811                             & 0.845                             & 0.864                             & 5.393                             & 0.029                              \\ 
			\cline{5-12}
			&                                          &                         &                                                                   & Y                   & 0.941                             & 0.900                             & 0.812                             & 0.854                             & 0.850                             & 5.136                             & 0.027                              \\ 
			\cline{2-12}
			& \multirow{2}{1in}{Attention U-Net + vgg16} & \multirow{2}{*}{31.9M}  & \multirow{2}{*}{45.32}                                            & N                   & 0.860                             & 0.877                             & 0.752                             & 0.810                             & 0.827                             & 5.215                             & 0.027                              \\ 
			\cline{5-12}
			&                                          &                         &                                                                   & Y                   & 0.946                             & 0.901                             & 0.808                             & 0.852                             & 0.860                             & 5.010                             & 0.025                              \\ 
			\cline{2-12}
			& \multirow{2}{1in}{Dense U-Net + vgg16}     & \multirow{2}{*}{20.2M}  & \multirow{2}{*}{41.54}                                            & N                   & 0.883                             & 0.881                             & 0.787                             & 0.831                             & 0.811                             & 5.082                             & 0.032                              \\ 
			\cline{5-12}
			&                                          &                         &                                                                   & Y                   & 0.960                             & 0.914                             & 0.820                             & 0.864                             & 0.880                             & 4.840                             & 0.023                              \\ 
			\cline{2-12}
			& \multirow{2}{1in}{DLA + vgg16}             & \multirow{2}{*}{23.7M}  & \multirow{2}{*}{44.21}                                            & N                   & 0.880                             & 0.859                             & 0.812                             & 0.835                             & 0.839                             & 5.142                             & 0.028                              \\ 
			\cline{5-12}
			&                                          &                         &                                                                   & Y                   & 0.968                             & 0.910                             & 0.820                             & 0.863                             & 0.890                             & 5.082                             & 0.024                              \\ 
			\cline{2-12}
			& \multirow{2}{1in}{RCA-IUnet (Ours)}        & \multirow{2}{*}{2.9M}   & \multirow{2}{*}{18.74}                                            & N                   & \textcolor{blue}{ \textbf{0.969}} & \textcolor{blue}{ \textbf{0.938}} & \textcolor{blue}{ \textbf{0.889}} & \textcolor{blue}{ \textbf{0.913}} & \textcolor{blue}{ \textbf{0.888}} & \textcolor{blue}{ \textbf{4.810}} & \textcolor{blue}{ \textbf{0.022}}  \\ 
			\cline{5-12}
			&                                          &                         &                                                                   & Y                   & \textbf{0.970}                    & \textbf{0.940}                    & \textbf{0.890}                    & \textbf{0.914}                    & \textbf{0.899}                    & \textbf{4.710}                    & \textbf{0.020}                     \\
			\hline
		\end{tabular}}
	\end{table*}
	
	\begin{figure}[b!]
		\centering
		\includegraphics[width=\columnwidth] {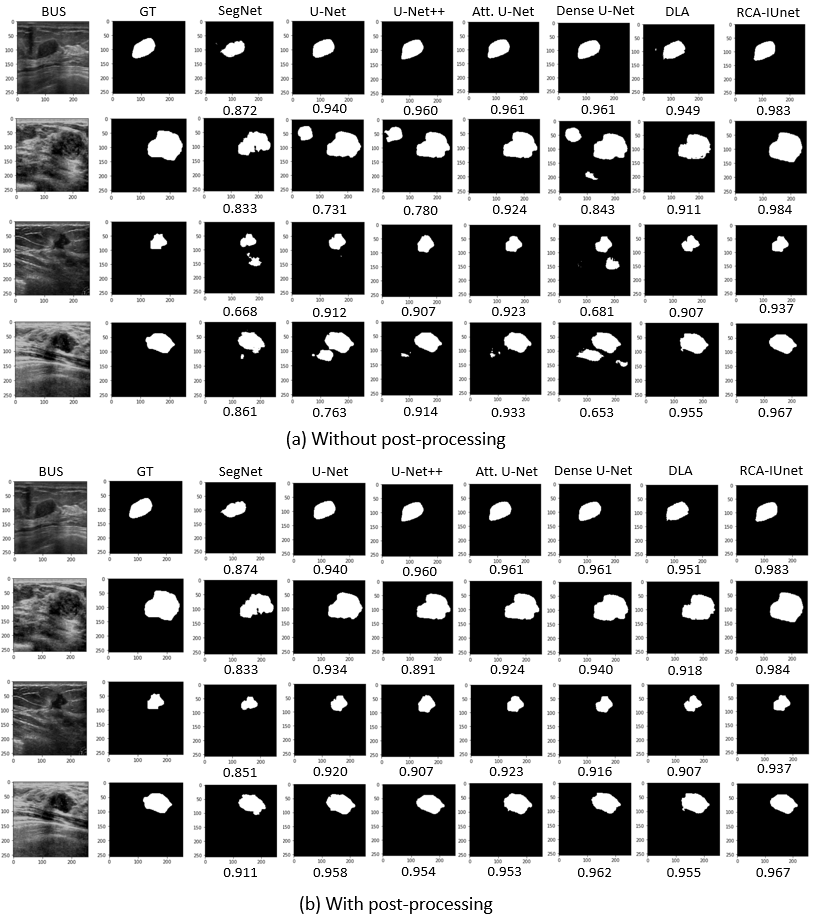}
		\caption{Qualitative comparison of BUS tumor segmentation results of the models: SegNet, U-Net, U-Net++, attention U-Net, dense U-Net, deep layer aggregation and RCA-IUnet, a) Without the post-processing and b) With the post-processing. The quantities indicate the dice score for each predicted mask.}
		\label{fig10}
	\end{figure}
	
	\section{Results and discussion}
\begin{figure}[H]
		\centering
		\includegraphics[width=\columnwidth] {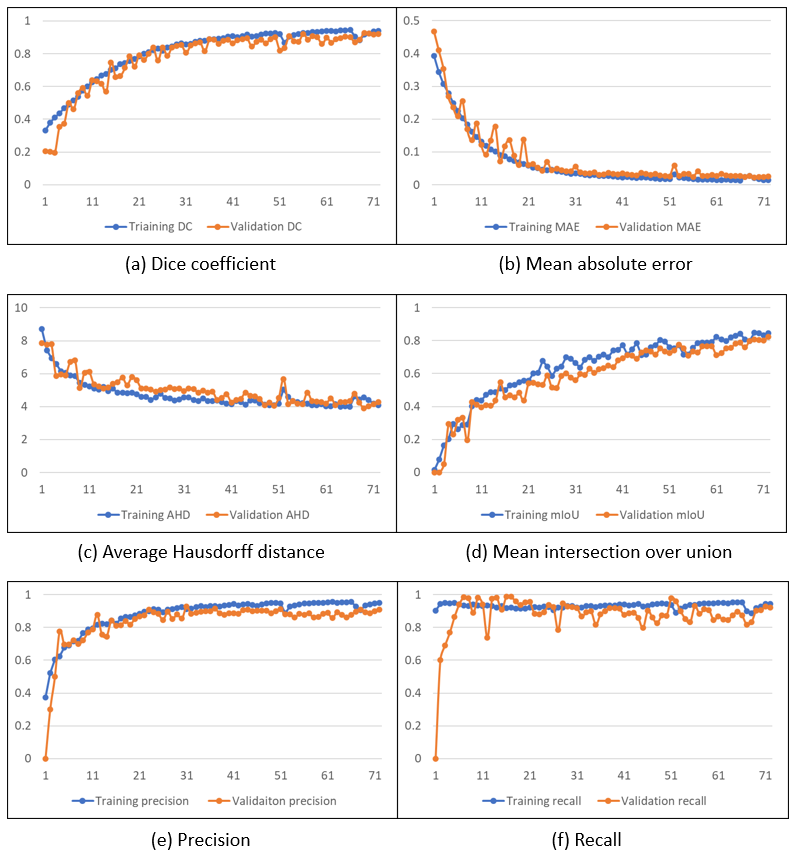}
		\caption{Summary of average training and validation scores: a) Dice coefficient, b) Mean absolute error, c) Average Hausdorff distance, d) Mean intersection over union, e) Precision and f) Recall, of RCA-IUnet model over BUSIS and BUSI datasets.}
		\label{fig11}
	\end{figure}

	The models produce a binary tumor segmentation mask for a given BUS image. The qualitative results of all the models with and without the post-processing are shown in Fig. \ref{fig10}. The generated segmentation mask along with the dice scores confirms the better performance of the RCA-IUnet model over other segmentation models. Fig. \ref{fig11} presents the mean segmentation performance of the RCA-IUnet model over the training and validation sets from both the datasets monitored during the training phase. From Fig. \ref{fig11}, it can be observed that the training and validation scores are promising and close to each other indicating that the RCA-IUnet model neither overfits nor underfits the training data and hence generates better segmentation masks.
	
	\begin{table*}
		\centering
		\caption{Ablation study of RCA-IUnet model. The best results and proposed model are highlighted in bold. RIC: residual inception convolution, HP: hybrid pooling and CSA: cross spatial attention, RCA-IUnet: U-Net + RIC + HP + CSA.}
		\label{tab3}
		\begin{tabular}{|l|P{1.6in}|l|l|l|l|l|l|l|l|} 
			\hline
			\multirow{2}{*}{Dataset} & \multirow{2}{*}{Model}             & \multirow{2}{*}{IT (ms)} & \multicolumn{7}{c|}{Tumor segmentation}                                                                                  \\ 
			\cline{4-10}
			&                                    &                                 & Acc. $\uparrow$ & Pr. $\uparrow$ & R $\uparrow$ & DC $\uparrow$ & mIoU $\uparrow$ & AHD $\downarrow$ & MAE $\downarrow$  \\ 
			\hline
			\multicolumn{1}{|c|}{\multirow{7}{*}{\rotatebox[origin=c]{90}{BUSIS}}}   & U-Net                              & 3.187                           & 0.680           & 0.521          & 0.553        & 0.536         & 0.527           & 6.433            & 0.095             \\ 
			\cline{2-10}
			& U-Net + RIC                        & 18.06                           & 0.920           & 0.881          & 0.864        & 0.872         & 0.869           & 5.001            & 0.022             \\ 
			\cline{2-10}
			& U-Net + CSA                        & 14.28                           & 0.911           & 0.873          & 0.860        & 0.866         & 0.862           & 5.120            & 0.022             \\ 
			\cline{2-10}
			& U-Net + RIC +   HP                 & 18.11                           & 0.930           & 0.901          & 0.884        & 0.892         & 0.883           & 4.701            & 0.021             \\ 
			\cline{2-10}
			& U-Net + CSA + HP                   & 14.93                           & 0.933           & 0.911          & 0.884        & 0.893         & 0.883           & 4.700            & 0.021             \\ 
			\cline{2-10}
			& U-Net + CSA + RIC                  & 18.31                           & 0.987           & 0.926          & 0.912        & 0.919         & 0.897           & 4.644            & 0.019             \\ 
			\cline{2-10}
			& U-Net + RIC + HP + CSA (RCA-IUnet) & 18.75                           & 0.990           & 0.954          & 0.920        & 0.937         & 0.910           & 4.632            & 0.019             \\ 
			\hline
			\multicolumn{1}{|c|}{\multirow{7}{*}{\rotatebox[origin=c]{90}{BUSI}}}    & U-Net                              & 3.185                           & 0.621           & 0.468          & 0.519        & 0.492         & 0.483           & 6.501            & 0.095             \\ 
			\cline{2-10}
			& U-Net + RIC                        & 18.06                           & 0.899           & 0.861          & 0.849        & 0.855         & 0.843           & 5.110            & 0.024             \\ 
			\cline{2-10}
			& U-Net + CSA                        & 14.28                           & 0.885           & 0.860          & 0.845        & 0.852         & 0.839           & 5.121            & 0.024             \\ 
			\cline{2-10}
			& U-Net + RIC +   HP                 & 18.11                           & 0.933           & 0.911          & 0.866        & 0.888         & 0.858           & 4.823            & 0.021             \\ 
			\cline{2-10}
			& U-Net + CSA + HP                   & 14.93                           & 0.923           & 0.920          & 0.861        & 0.890         & 0.860           & 4.813            & 0.021             \\ 
			\cline{2-10}
			& U-Net + CSA + RIC                  & 18.31                           & 0.968           & 0.932          & 0.879        & 0.905         & 0.889           & 4.751            & 0.020             \\ 
			\cline{2-10}
			& U-Net + RIC + HP + CSA (RCA-IUnet) & 18.74                           & 0.970           & 0.940          & 0.891        & 0.914         & 0.899           & 4.710            & 0.020             \\
			\hline
		\end{tabular}
	\end{table*}
	
	\begin{table*}[h!]
		\centering
		\caption{Cross data validation of RCA-IUnet model with fine tuning. $D_1$ - BUSIS dataset, $D_2$ - BUSI dataset. Scenario $D_1 - D_2$ indicate model is trained on $D_1$, fine-tuned and tested on $D_2$, whereas vice-versa for scenario $D_2 - D_1$.}
		\label{tab4}
		\begin{tabular}{|l|l|l|l|l|l|l|l|}
			\hline
			Scenario  & Acc. $\uparrow$ & Pr. $\uparrow$ & R $\uparrow$   & DC $\uparrow$  & mIoU $\uparrow$ & AHD $\downarrow$ & MAE $\downarrow$  \\  \hline
			$D_1 - D_2$ & 0.957 & 0.913 & 0.885 & 0.901 & 0.855 & 4.879 & 0.023 \\ \hline
			$D_2 - D_1$ & 0.990 & 0.959 & 0.921 & 0.936 & 0.926 & 4.897 & 0.019 \\ \hline
		\end{tabular}
	\end{table*}
	
	It is also observed that among the tested models, the post-processing techniques have minimal impact on the performance of the RCA-IUnet model, indicating that the model produces a segmentation mask with very low false positive and false negative predictions of the tumor regions. However, there is a noticeable improvement in the performance of other models by using post-processing, indicating that these models generate high false predictions and hence relies on further refinement to improve the results. For instance, in Fig.~\ref{fig10}, the segmentation mask generated for the second sample by U-Net without and with post-processing has dice scores of $0.731$ and $0.934$ respectively, while the RCA-IUnet model produces same results with a better dice score of $0.984$. Besides, the overall quantitative results are shown in Table \ref{tab2} along with the comparative analysis with other state-of-the-art models in terms of evaluati-
	\noindent on metrics described in Table \ref{tab1}. The proposed model outperformed with best segmentation scores and minimal inference time while having considerably less number of training parameters. 
	
	The effectiveness of each proposed component of the RCA-IUnet model is analyzed in Table \ref{tab3}. This ablation study is conducted by adding the proposed components to base U-Net model. Here U-Net is a skeleton model of complete RCA-IUnet model that consists of default depth-wise separable convolutions, max pooling operations and skip connections with four stages of encoding and decoding. This study is conducted with the same training, validation and testing sets of both datasets over various combinations to form different models by adding components to the U-Net model such as U-Net + CSA, U-Net + RIC + HP, etc. The performance of each model is compared using segmentation metrics along with the inference time (IT). From Table \ref{tab3} it can be inferred that RIC and CSA are core components that derive the outperforming nature of the RCA-IUnet model as shown for models: U-Net + RIC, U-Net + CSA and U-Net + RIC + CSA. The residual inception convolution enables the network to capture multi-scale feature representation, and cross-spatial attention enables the network to draw attention towards the most relevant features. As compared to max pooling, hybrid pooling plays a vital role with efficient downsampling to further improve the results as shown for the models: U-Net + RIC + HP vs U-Net + RIC and U-Net + CSA + HP vs U-Net + CSA. With the achieved quantitative results it is evident that each component contributes to improving the segmentation performance of the RCA-IUnet model. Though this segmentation performance is delivered with increased inference time as compared to the base U-Net model but is comparatively lesser as compared to the existing models as shown in Table \ref{tab2}.
	
	To further establish the robustness of the proposed model a cross data validation is performed as shown in Table \ref{tab4}. The testing is performed with two scenarios: 1) model pre-trained on BUSIS dataset is tested on BUSI dataset, and 2) model pre-trained on BUSI dataset and is tested on BUSIS dataset, by fine-tuning. The model achieved similar results as highlighted in Table \ref{tab2} and Table \ref{tab3}. This indicates that the proposed model can adapt to a new dataset by just fine-tuning without compromising the performance.	
	
	\section{Conclusion}
	This article proposes a deep learning based model, residual cross-spatial attention inception U-Net (RCA-IUnet), for breast tumor segmentation in ultrasound imaging. The RCA-IUnet model is designed with a state-of-the-art U-Net model that uses residual inception depth-wise separable convolution and hybrid pooling (max pooling and spectral pooling) layers along with the cross-spatial attention filter in the long skip connections to better propagate and extract the feature maps concerning the tumor region. With exhaustive trials, the proposed model achieved significant improvement over the state-of-the-art models with minimal training parameters and inference time on two publicly available datasets to generate tumor segmentation mask. Moreover, the ablation study describes the significance of each component of the model towards tumor segmentation, where residual inception convolution (RIC) and cross-spatial attention (CSA) components displayed a major contribution in the achieved results. As an extension, the attention component could further be improved by incorporating a channel attention filter to focus on most relevant feature layers. Overall the performance of the model could further be improved by incorporating deeper feature extraction layers, hybrid or ensemble learning leading towards better feature representation for tumor regions. Besides, the scope of this model is not limited to tumor segmentation in breast ultrasound imaging, it can also provide potentially useful results with other modalities for biomedical image segmentation. 
	
	\section*{Acknowledgment}
	We thank our institute, Indian Institute of Information Technology Allahabad (IIITA), India and Big Data Analytics (BDA) lab for allocating the centralised computing facility and other necessary resources to perform this research. We extend our thanks to our colleagues for their valuable guidance and suggestions.
	
	\bibliographystyle{elsarticle-num}
	\bibliography{reference}
	
\end{document}